\documentclass[a4paper,10pt]{article}
\usepackage{amsmath}
\usepackage{color}
\usepackage{amssymb}

\title{Should quantum theory change in the light of gravity?}
\author{Johan Noldus}
\begin{document}
\maketitle
\begin{abstract}
In this paper, we study implications of the geometrical nature of space-time for some of the basic tenets of quantum mechanics. That is, we study two different implications of the principle of general covariance; first we quantize a reparametrization invariant theory, the free particle in
Minkowski spacetime and point out in detail where this theory fails (no-
tably these comments appear to be missing in the literature). Second we
study the covariance of quantum field theory and show how it connects
to causality, the outcome of this study is that QFT is what we shall call
ultra weakly covariant with respect to the background spacetime. Third,
we treat the question of whether evolution in quantum theory (apart from
the measurement act) needs to be unitary, it is easily shown that a per-
fectly satisfying probabilistic interpretation exists which does not require
unitary evolution. Fourth, we speculate on some modifications quantum
theory should undergo in order for it to be generally covariant. The results in this paper hint at a profound change of the theory in which causality as a fundamental principle is abandonned.  
\end{abstract}
\section{Introduction}
In this paper, we study in detail the extend to which the principles underlying
quantum theory and general relativity agree or disagree. Since different people
mean different things when speaking about general covariance \cite{Wald}, it is perhaps
good to name them differently: strong covariance means that the equations of
motion do not depend upon the coordinate system at hand and moreover, all
fields involved are dynamical\footnote{Meaning that they satisfy nontrivial equations of motion.},
 weak covariance means that all predictions of the
theory do not depend upon the coordinate system at hand, but some fields are
background fields and ultra weak covariance indicates that the predictions of the
theory remain invariant for coordinate systems defining foliations of spacetime
which are spacelike with respect to the background structure. As we shall prove,
quantum field theory turns out to be ultra weakly but not strongly covariant;
moreover, ultra weak covariance is all we need to ensure classical causality, that
is spacelike separated field operators do (anti) commute with each other. It is argued that progress in 
quantum theory can be made if we know how to reformulate quantum field theory into a weakly covariant fashion.
\section{Covariant quantization of the free relativistic particle.}
Henceforth the metric structure on spacetime will have signature $(+ - - -)$ and
the action principle at hand is given by
$$\mathcal{S} = m \int_{\tau_0}^{\tau_1} \sqrt{\eta_{\mu \nu} \dot{x}^{\mu} \dot{x}^{\nu}} d\tau.$$
The canonical momenta $p_{\mu}$ are given by
$$p_{\mu} = m \frac{\eta_{\mu \nu} \dot{x}^{\nu}}{\sqrt{\eta_{\alpha \beta} \dot{x}^{\alpha} \dot{x}^{\beta}}}$$
and satisfy the constraint
                                 $$ \eta_{\mu \nu} p_{\mu} p_{\nu} = m^2.$$
Denoting by $v(\tau) = \sqrt{\eta_{\mu \nu} \dot{x}^{\mu} \dot{x}^{\nu}}$, the Legendre transformation is given by                    
$$\dot{x}^{\mu} = \frac{v(\tau) \eta^{\mu \nu}p_{\nu}}{m}$$
and a\footnote{There exist several Hamiltonians, for example $\mathcal{H}(\tau) = v(\tau) \left( \frac{1}{m} \eta^{\mu \nu}p_{\mu}p_{\nu}
- m \sqrt{\eta^{\mu \nu}p_{\mu}p_{\nu}}\right)$ is another one.} Hamiltonian giving the correct equations of motion is given by
                                $$\mathcal{H}(\tau) = \frac{1}{2m}v(\tau) \left( \eta^{\mu \nu}p_{\mu}p_{\nu} - m^2 \right).$$
Clasically, we have that $\dot{p}_{\mu}(\tau) = 0$ and
            $$x^{\mu}(\tau) = x^{\mu}_{0} + \frac{\eta^{\mu \nu} p_{\nu}}{m} \int_{\tau_0}^{\tau}v(s)ds$$               ÃƒÆ’Ã†â€™Ãƒâ€¦Ã¢â‚¬â„¢ÃƒÆ’Ã‚Â¢ÃƒÂ¢Ã¢â‚¬Å¡Ã‚Â¬Ãƒâ€šÃ‚Â¡
meaning that the $x^{\mu}(\tau)$ are not predictions of the theory (since they depend on
$v$), but nevertheless observables and they are called partial observables by some
authors.  What we do know however is that for $p_0 > 0$, $t(\tau)$ is an increasing
function of $\tau$ and for suitable $v$ the entire $t$ range will be covered, hence we can
ask the question, what is $x^i(t)$?  It is given by
$$x^i(t) = x^{i}_{0} + \frac{p_i}{p_0}(t - t_0)$$                         
and is a prediction of the theory since it does not depend upon $v$. Therefore,
the theory forcasts the spatial coordinates at some fixed time coordinate $t$, 
but the statement doesnÃƒÆ’Ã†â€™Ãƒâ€šÃ‚Â¢ÃƒÆ’Ã‚Â¢ÃƒÂ¢Ã¢â€šÂ¬Ã…Â¡Ãƒâ€šÃ‚Â¬ÃƒÆ’Ã‚Â¢ÃƒÂ¢Ã¢â€šÂ¬Ã…Â¾Ãƒâ€šÃ‚Â¢t involve $\tau$ in any way.  Now we come to a delicate
point, the right hand side of the above expression is nonlocal in time, it involves
dynamical quantities evaluated at times $\tau_0$ and $\tau$, and therefore it is not a Dirac
observable\footnote{For instance, we don't know how to evaluate Poisson brackets of the type $\{a(ÃƒÆ’Ã†â€™Ãƒâ€šÃ‚ÂÃƒÆ’Ã‚Â¢ÃƒÂ¢Ã¢â‚¬Å¡Ã‚Â¬Ãƒâ€¦Ã‚Â¾), b(\widetilde{\tau}) \}$.  
This is different in quantum mechanics where we can commute operators at different times.}.  
Since the constraint $\mathcal{C}(\tau) = \eta^{\mu \nu} p_{\mu}(\tau) p_{\nu}(\tau) - m^2$ is first class, it 
is a generator of gauge transformations, moreover $\dot{\mathcal{C}}(\tau) = 0$ weakly\footnote{This is precisely so in gravity.}. Since
$x^{\mu}(\tau) + \frac{p_{\mu}(\tau)}{p_{0}(\tau)}t(\tau)$ is a Dirac observable of the theory, the Poisson bracket with
$\mathcal{C}(\tau)$ should vanish. Indeed,
\begin{eqnarray*}
\{ x^{\mu}(\tau) + \frac{p_{\mu}(\tau)}{p_{0}(\tau)}t(\tau), \mathcal{C}(\tau) \} & = & \{ x^{\mu}(\tau), - p_{\mu}^{2}(\tau) \} + 
\frac{p_{\mu}(\tau)}{p_{0}(\tau)} \{ t(\tau), p_{0}^{2}(\tau) \} \\*
& = & - 2p_{\mu}(\tau) + 2 \frac{p_{\mu}(\tau)}{p_{0}(\tau)}p_{0}(\tau) \\*
& = & 0.              
\end{eqnarray*}                        
However, to measure this observable (constant of motion) we need four different
measurements; classically this is fine, but what is the meaning quantum me-
chanically?  At any rate, classically, it is generically the case that there is a huge
sensitivity of the number of nontrivial observables (not Dirac observables) on
the initial conditions.  If one thinks about general relativity, one has spacetimes
where one can find global coordinate charts formed by curvature invariants (and
invariants not included in this set are predicted from these coordinates) while for
example in Minkowski spacetime, no local statements based upon these invari-
ants can be made. In what follows about the quantum theory, our comments
shall be focused around the lack of localizability, the scalar product, and the
problem of time. The momentum operators are constant and can for all times
$\tau$ be given by the operators
                                         $$p_{\mu} = -i\partial_{\mu} $$
while the $x^{\mu}(\tau_{0})$ are given by
                                         $$x^{\mu}(\tau_{0}) = \widehat{x}^{\mu}$$
where $\widehat{x}^{\mu}$ is the multiplication operator with $x^{\mu}$.  $\widehat{x}^{\mu}(\tau)$ is then given by
$$\widehat{x}^{\mu}(\tau) = \widehat{x}^{\mu} - \frac{i\hbar \eta^{\mu \nu}\partial_{\nu}}{m} \int_{\tau_{0}}^{\tau}v(s)ds.$$
These operators have a unique Self-Adjoint extension on $L^2 (\mathbb{R}^4 , d^4 x)$ and the
constraint operator is given by
                                    $$\hbar^2 \eta^{\mu \nu}\partial_{\mu}\partial_{\nu} + m^2 $$
Physical states $\Psi$ must then satisfy
                                   $$\hbar^2 \eta^{\mu \nu}\partial_{\mu}\partial_{\nu}\Psi + m^2\Psi = 0.$$
A first problem one encounters is that there exist no normalizable physical
states, in the literature one proposes to consider the Klein-Gordon scalar prod-
uct
                     $$(\Phi, \Psi) = \int d^3 x \left( i\partial_t \Phi^{\star} \Psi - i\Phi^{\star} \partial_t  \Psi \right)_{|t=0}$$
on the subspace of positive frequency solutions. However, the time operator $\widehat{t}$ is
not Hermitian with respect to this scalar product! This problem is circumvented
by considering the time independent scalar product
                          $$ (\Phi, \Psi) = \int d^3 x \Phi^{\star} \Psi_{|t=0}$$
on the same subspace.  Let us now treat the problem of time; the time evolution
operator $U (\tau)$ is given by
                $$U(\tau) = e^{ -\frac{i}{m \hbar} \int_{\tau_0}^{\tau} v(s)ds
\left( \hbar^2 \eta^{\mu \nu}\partial_{\mu}\partial_{\nu} + m^2  \right)} $$                      
and
                               $$\widehat{x}^{\mu}(\tau) = U (ÃƒÆ’Ã†â€™Ãƒâ€šÃ‚Â¢ÃƒÆ’Ã¢â‚¬Â¹ÃƒÂ¢Ã¢â€šÂ¬ ÃƒÆ’Ã‚Â¢ÃƒÂ¢Ã¢â‚¬Å¡Ã‚Â¬ÃƒÂ¢Ã¢â‚¬Å¾Ã‚Â¢ \tau)\widehat{x}^{\mu}U (\tau).$$
Now, there happens something funny, it is so that
                     $$(U(\tau)\Psi, \widehat{x}^{\mu}U(\tau)\Phi) \neq (\Psi, U(ÃƒÆ’Ã†â€™Ãƒâ€šÃ‚Â¢ÃƒÆ’Ã¢â‚¬Â¹ÃƒÂ¢Ã¢â€šÂ¬ ÃƒÆ’Ã‚Â¢ÃƒÂ¢Ã¢â‚¬Å¡Ã‚Â¬ÃƒÂ¢Ã¢â‚¬Å¾Ã‚Â¢\tau)\widehat{x}^{\mu}U(\tau)\Phi)$$
and the reason is that the operator $ÃƒÆ’Ã†â€™Ãƒâ€šÃ‚Â¢ÃƒÆ’Ã¢â‚¬Â¹ÃƒÂ¢Ã¢â€šÂ¬ ÃƒÆ’Ã‚Â¢ÃƒÂ¢Ã¢â‚¬Å¡Ã‚Â¬ÃƒÂ¢Ã¢â‚¬Å¾Ã‚Â¢i \partial_t$ is only Hermitian when both arguments in the scalar product are positive frequency physical states.  The operator
$\widehat{x}^{\mu}$ however, maps physical states to nonphysical states and hence the Hermitic-
ity property does not apply.  This means that there is a disagreement between
the Schroedinger picture and the Heisenberg picture; the former has no time
evolution of physical matrix elements, while the latter has! So, the problem of
time appears to be absent in the Heisenberg picture. Now, we come to the most
serious problem, the lack of localization: the operators $\widehat{x}^{\mu}(\tau)$ do not commute
with the constraint operator and hence both cannot be simultaneously diagonalized. 
 Therefore, it is not possible to make sharp measurements of space and time
according to Von Neumann measurement theory. Funny enough, the expectation values of (powers of) position operators $\widehat{x}^{\mu}$ indicate localization (due to the
fact that the scalar product takes one time slice, expectation values of operators
do depend upon the slice). For example $(\Psi, t^n \Psi)$ = 0 hence $\delta t = 0$, however this
perfect localization fails for $t(\tau), \tau \neq \tau_0$, and realistic wave packages\footnote{Therefore it is not fair to say that time is fixed to some definite value.  
Furthermore $\delta t(\tau) = a \int_{\tau_{0}}^{\tau} v(s)ds$ with $a > 0$ with average $\langle t(\tau) \rangle = b \int_{\tau_{0}}^{\tau}
v(s)ds$, $b > 0$.  One could at each stage where $\delta t$ exceeds some treshold value, replace the state $\Psi$ by one which is more
localized in time (smaller $a$). But then it appears to become impossible to localize the wavepackage in space.} ! Since
von Neumann measurement theory fails, I think it is fair to say that the above
quantization is inadequate and therefore one might conclude that quantum mechanics, 
with its current measurement theory, is not strongly covariant. Similar problems should be expected for quantum gravity;
 actually it turns out that the situation gets even worse. Although it is possible to (formally) construct an
appropriate Klein Gordon like scalar product on superspace, it turns out to be
impossible to find a subspace of positive frequency solutions with positive norm
\cite{Isham}. This has lead to the suggestion of third quantization.
\section{Ultra weak covariance of quantum field theory.}
What follows is fairly obvious but it is nevertheless good do at least once in
your lifetime. We are going to prove that quantum field theory is ultra weakly
covariant and that therefore causality always holds. For sake of simplicity we
shall argue from the viewpoint of free massive Klein Gordon theory given by
the action
$$S = \frac{1}{2}\int d^4 \, \sqrt{-g} \left( \eta^{\mu \nu}\partial_{\mu} \psi \partial_{\nu} \psi - m^2 \psi^2  \right)$$
in a general curved spacetime. Let $(t,x^i)$ be a foliation of spacetime; note first
that Hamiltonian evolution is only well defined if the hypersurfaces of constant
$t$ do not become null. Indeed, $g^{tt}$ is proportional to the determinant of $g_{ij}$ which
vanishes if and only if the $\partial_i$ span a null hypersurface.  A vanishing $g^{tt}$ means
that the Legendre transform and therefore the Hamiltonian becomes singular, 
which effectively stops evolution.  Consider two foliations $(t,x^i)$ and $(\widetilde{t},\widetilde{x}^i)$ of
spacetime with spacelike leaves which coincide for $t \in \left[t_0 , t_0 + \epsilon \right]$.  Choosing
intial operators on some Hilbert space $\mathcal{H}$ so that the intial value conditions
\begin{eqnarray*}
\left[ \psi(t_0,x^i), \psi(t_0,y^i) \right] & = & 0 \\*
\left[ \psi(t_0,x^i), \pi(t_0,y^i) \right] & = & i \delta(x^i - y^i) \\*
\left[ \pi(t_0,x^i), \pi(t_0,y^i) \right] & = & 0   
\end{eqnarray*}
are satisfied, covariance of the equations of motion guarantees that the solutions
will be the same and in particular one has that
$$\left[ \psi(t(\widetilde{t},\widetilde{x}^j),x^i(\widetilde{t},\widetilde{x}^j)), \psi(t(\widetilde{t},\widetilde{y}^j),x^i
(\widetilde{t},\widetilde{y}^j)) \right]  =  0.$$                  
One can get rid of the common initial value tube by repeating the same argument 
for $(\widetilde{t}, \widetilde{x}^i)$ ``backwards'' in time.  It is now easy to see that the previous (initial
value) conditions hold for any coordinate system with spacelike leaves. This
shows that the \textit{solution} $\psi(t, x^i)$ behaves in a covariant fashion with respect
to spacelike foliations.  This does not show of course that were one to choose
another initial representation on $\mathcal{H}$, that both theories are equivalent in some
sense; it is well known that inequivalent representations of the CCR algebra
exist.  But what we have shown here is that any of these theories behaves
covariantly and hence causality is respected.  It is perhaps illuminating to show
this explicitely for massive Klein Gordon theory in flat Minkowski space time,
since the calculation reveals the importance of spacelike folations (it doesn't
work for timelike ones). This is left as a not entirely trivial exercise for the
reader.
\section{Unitarity and quantum mechanics}
It is often repeated that quantum mechanics needs to be unitary because one
needs a coherent probability interpretation. What we shall show here is that one
can consistently define experimental probabilities as outcomes of measurements
without reference to unitary evolution at all. This is accomplished in the Heisen-
berg picture. Let $A$ denote an observable at time $t = 0$ and $U(t) = e^{iHt}$ be
the unitary evolution operator.  Clearly the spectrum of $A$ is the same as that of
$A(t) = U(-t)AU(t)$ and the orthogonal projector on the subspace of eigenvalue
$\lambda$ evolves as
                        $$P_{\lambda}^{A}(t) = U(-t)P_{\lambda}^A(0)U(t).$$
Denote by $\Psi$ the initial state, then the probability that at time $t$ we measure
for the observable $A$ the eigenvalue $\lambda$ is given by
                $$\textrm{Prob}(A,\lambda,t) = \frac{\left( \Psi, P_{\lambda}^{A}(t) \Psi \right)}{\left( \Psi, \Psi \right)}.$$
This can be extended to an arbitrary number of arguments: the probability that
$O$ is measured to be $\mu$ at time $s$ given that $A$ was measured to be $\lambda$ at time $t$
is given by
$$\textrm{Prob}(O,\mu,s |A,\lambda,t) = \frac{\left( \Psi,  P_{\lambda}^{A}(t) P_{\mu}^{O}(s) P_{\lambda}^{A}(t) \Psi \right)}{\left(
 \Psi,  P_{\lambda}^{A}(t) \Psi \right)}.$$
One simply notices that all above probabilities are (a) well defined in the sense
that they sum up to one if all alternatives are taken into account and (b) their
expression does not require the evolution operator $U(t)$, all that is needed are
time dependent operators. Of course unitarity serves also other purposes, one of
which we spoke about before, such as the preservation of the commutation rela-
tions in time. So, one needs to study what will happen to those in a nonunitary
quantum theory.
\section{Conclusions and outlook.}
We have learned that the Schroedinger and Heisenberg picture are two inequiv-
alent theories and that the latter is the only plausible one. However, it is not
sufficient to simply adapt the Heisenberg picture, the Von Neumann theory of
measurement is not valid anymore and a substitute for an adquate probabilistic
interpretation needs to be found. Of course, it may just be that something
else is wrong at a more fundamental level. One way to make progress would
be to redefine quantum field theory in a weakly covariant fashion, to make it
really four dimensional as to speak. Certainly such endeavour has already been
undertaken by the algebraic formulation of free quantum field theory, but this
appears not to be sufficient to me. One notices that the commutation relations
are of singular nature leading to distribution valued operators. This would lead
to serious difficulties if all measurements were assumed to be dynamical (ie,
Dirac observables) and in section two we have already dismissed such point of
view. We have shown that the CCR algebra is compatible with ultra weak
covariance; however, the computations suggested at the end of section three
show that they do not in a weakly covariant framework. Therefore, we have
to look for a completely new way of devising a quantum theory in which the
role of causality is an emergent and not a fundamental property: measurement
influences should be allowed to tunnel moderately through the light cone as one
would expect in a full theory of quantum gravity\footnote{The same conclusion can be reached from a different perspective, see \cite{Isham}.}. It could be that if these
issues were resolved, some Von Neumann type measurement rule would persist
and therefore, our first priority consists in understanding the emergence of the
(approximate) canonical (anti)commutation relations.

\end{document}